\documentclass[aip, reprint, twocolumn, a4paper]{revtex4-1}
\pdfoutput=1
\usepackage{graphicx}
\usepackage{amsfonts}
\usepackage{amsmath}

\begin{document}
	
\begin{abstract}
Microfocused Brillouin light scattering (BLS) and microwave absorption (MA) are used to study magnon-photon coupling in a system consisting of a split-ring microwave resonator and a yttrium iron garnet (YIG) film. The split-ring resonantor is defined by optical lithography and loaded with a 1\,$\mu$m-thick YIG film grown by liquid phase epitaxy. BLS and MA spectra of the hybrid system are simultaneously recorded as a function of the applied magnetic field magnitude and microwave excitation frequency. Strong coupling of the magnon and photon modes is found with a coupling strength of $g_\text{eff}/2 \pi = 63$\,MHz. The combined BLS and MA data allows to study the continuous transition of the hybridized modes from a purely magnonic to a purely photonic mode by varying the applied magnetic field and microwave frequency. Furthermore, the BLS data represent an up-conversion of the microwave frequency coupling to optical frequencies.
\end{abstract}

\title{Combined Brillouin light scattering and microwave absorption study of magnon-photon coupling in a split-ring resonator/YIG film system}

\author{S.~Klingler}
\email{stefan.klingler@wmi.badw.de}
\author{H.~Maier-Flaig}
\affiliation{Walther-Mei{\ss}ner-Institut, Bayerische Akademie der Wissenschaften, Walther-Mei{\ss}ner-Stra{\ss}e 8, 85748 Garching, Germany}
\affiliation{Physik-Department, Technische Universit\"{a}t M\"{u}nchen, 85748 Garching, Germany}

\author{R.~Gross}
\affiliation{Walther-Mei{\ss}ner-Institut, Bayerische Akademie der Wissenschaften, Walther-Mei{\ss}ner-Stra{\ss}e 8, 85748 Garching, Germany}
\affiliation{Physik-Department, Technische Universit\"{a}t M\"{u}nchen, 85748 Garching, Germany}
\affiliation{Nanosystems Initiative Munich (NIM), 80799 Munich, Germany}

\author{C.-M.~Hu}
\affiliation{Department of Physics and Astronomy, University of Manitoba, Winnipeg, Manitoba R3T2N2, Canada}

\author{H.~Huebl}
\affiliation{Walther-Mei{\ss}ner-Institut, Bayerische Akademie der Wissenschaften, Walther-Mei{\ss}ner-Stra{\ss}e 8, 85748 Garching, Germany}
\affiliation{Physik-Department, Technische Universit\"{a}t M\"{u}nchen, 85748 Garching, Germany}
\affiliation{Nanosystems Initiative Munich (NIM), 80799 Munich, Germany}

\author{S.T.B.~Goennenwein}
\affiliation{Walther-Mei{\ss}ner-Institut, Bayerische Akademie der Wissenschaften, Walther-Mei{\ss}ner-Stra{\ss}e 8, 85748 Garching, Germany}
\affiliation{Physik-Department, Technische Universit\"{a}t M\"{u}nchen, 85748 Garching, Germany}
\affiliation{Nanosystems Initiative Munich (NIM), 80799 Munich, Germany}

\author{M.~Weiler}
\affiliation{Walther-Mei{\ss}ner-Institut, Bayerische Akademie der Wissenschaften, Walther-Mei{\ss}ner-Stra{\ss}e 8, 85748 Garching, Germany}
\affiliation{Physik-Department, Technische Universit\"{a}t M\"{u}nchen, 85748 Garching, Germany}

\maketitle

The interaction between light and magnetic matter is of long-standing and fundamental interest. In recent years, the coupling of elementary excitations of the light field (photons) to those of the spin system (magnons) regained interest due to potential applications in quantum information processing.\cite{Imamoglu2009, Wesenberg2009, Xiang2013}
For example, hybrid quantum system are discussed as potential candidates for the up- and down-conversion of quantum signals from the optical to the microwave domain and vice versa. One way of interfacing the microwave regime is to magnetically dipole couple the spin ensemble to a microwave resonator.\cite{Huebl2013,Zhang2014,Bai2015, Lambert2016, Maier-Flaig2016} The prerequisite for information transfer on the quantum level is to realize a large coupling strength exceeding the loss rates of the subsytems,\cite{Huebl2013, Tabuchi2015,Lambert2016} here, the microwave resonator and the spin ensemble. Since the coupling rate is proportional to the square root of the number of participating spins\cite{Huebl2013, Soykal2010} ferromagnets with a high spin density are ideal for the creation of strongly coupled, hybridized magnon-photon modes.\cite{Soykal2010, Soykal2010a} 

Recently, magnon-photon coupling has been investigated in several experiments where a microwave cavity was loaded with yttrium iron garnet (YIG) and the microwave transmission and/or reflection was measured as a function of the applied magnetic field.\cite{Huebl2013,Zhang2014,Bai2015, Lambert2016}
Furthermore, spin pumping in combination with the dc inverse spin Hall effect has been employed as a detection scheme for sensing the magnonic part of magnon-photon polaritons in magnetic thin film heterostructures,\cite{Bai2015, Maier-Flaig2016} which is effectively a down-conversion of the coupling to dc. 

Here, we report on the up-conversion of the strong coupling of photons in a micropatterned microwave split-ring resonator\cite{Bhoi2014,Gay-Balmaz2002} (SRR) and a YIG film to optical frequencies. In particular, we use Brillouin light scattering (BLS) spectroscopy and microwave absorption (MA) measurements to simultaneously probe both magnonic and photonic excitations in a coupled SRR/YIG film system, which is a first step towards wavelength up-conversion. We observe a clear mode anti-crossing indicating a hybridization of the magnon and microwave photon modes in the strong coupling regime. The transition from a dominantly photonic to a dominantly magnonic mode across the anti-crossing is investigated by evaluating the BLS and MA signal amplitudes. Furthermore, light-polarization dependent measurements give insight into the nature of inelastic scattering of optical-frequency photons by the hybridized mode of microwave-frequency magnons.

\begin{figure}[htb]%
	\begin{center}%
		\scalebox{1}{\includegraphics[width=0.9\linewidth,clip]{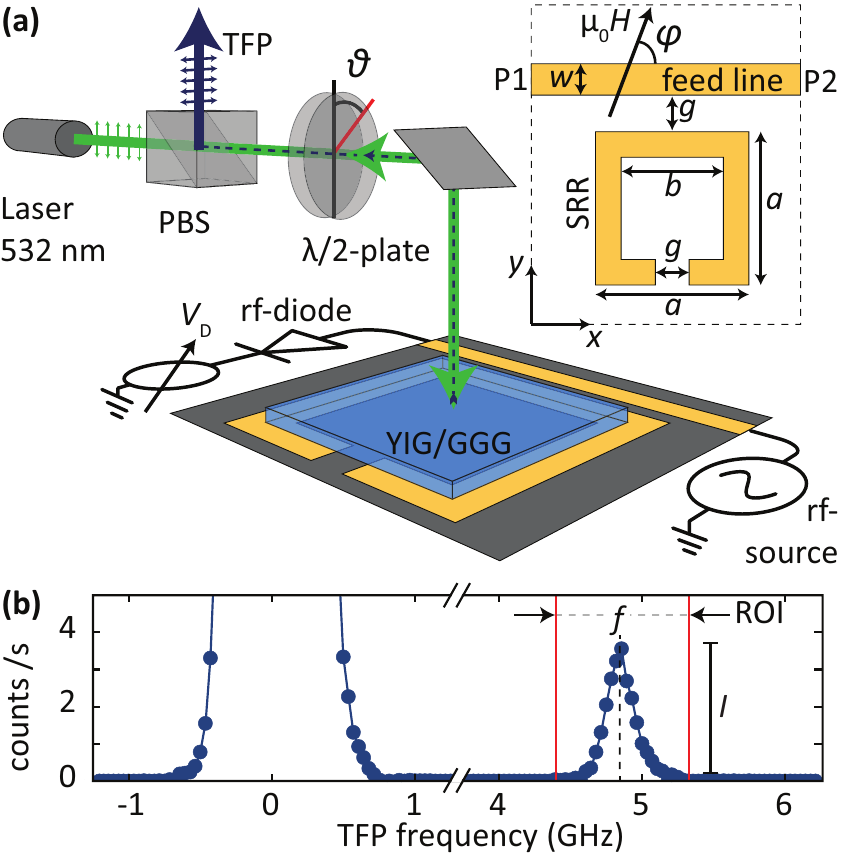}}%
	\end{center}%
	\caption{\label{setup} 
		(a) Experimental setup: A microwave signal with frequency $f$ is applied to the feedline which is inductively coupled to the SRR. The YIG film is centered onto the SRR. The microwave transmission through the feedline is detected with a diode and a voltmeter as $V_\text{D}$. A polarized laser beam passes a polarizing beam splitter (PBS) and a $\lambda/2$-plate and is focussed on the surface of the YIG film by a microscope objective lens (not shown). The backscattered light passes again the $\lambda/2$-plate and the PBS, before it reaches the Tandem-Fabry-P\'{e}rot interferometer (TFP). The polarization of the backscattered light is determined by changing $\vartheta$. Inset: Sketch of the SRR system. (b)~Typical BLS spectrum as a function of the detuning from the laser line, with the reference laser peak at 0\,GHz and the anti-Stokes signal at frequency $f$ obtained for 100 averages.
	}%
\end{figure}%

A sketch of the experimental setup is shown in Fig.~\ref{setup}\,(a). The setup consists of three parts: (i) the SRR/YIG film system where the coupled magnon-photon dynamics takes place, (ii) the microwave absorption setup to investigate the photonic excitations, and (iii) the BLS setup to analyze the magnonic excitations. 
The SRR system is fabricated by optical lithography on a 508\,$\mu$m-thick Rogers RT/duroid 5870  substrate with a double-sided 35\,$\mu$m copper coating. It consists of a $50\,\Omega$ impedance matched feedline with a width of $w=1.4$\,mm inductively coupled to the SRR.\cite{Bhoi2014, Gay-Balmaz2002} The square SRR has an outer edge length of $a=6.5$\,mm while the inner edge length is $b=3.5$\,mm. The gap width and distance to the feedline are both $g=0.2$\,mm. The YIG film has a lateral size of 5$\times$5\,mm$^2$ and a thickness of 1\,$\mu$m. It was grown by liquid phase epitaxy on a (111)-oriented Gadolinium Gallium Garnet (GGG) substrate. The YIG/GGG heterostructure is positioned with the substrate side down in the center of the SRR. In this way, the YIG film is optically accessible from above for the BLS measurements. The SRR can be described as an $LRC$-circuit\cite{Gay-Balmaz2002,Bhoi2014} which can absorb energy of an oscillating microwave field close to its resonance frequency $f_\text{SRR}=4.96$\,GHz. The rf-current in the SRR creates oscillating magnetic fields which drive the magnetization precession in the YIG film.\cite{Bhoi2014} 
An external magnetic field 40\,mT$\leq \mu_0 H \leq$200\,mT is applied in the film plane at an angle $\varphi=20^\circ$ relative to the feedline ($x$-axis). This angle is chosen to comply with geometrical restrictions of the BLS setup.

The microwave absorption is recorded by connecting port~1~(P1) and port~2~(P2) of the feedline to a microwave source and a microwave diode, respectively. The SRR is excited with microwave radiation in the frequency range $4.8\leq f \leq 5.2$\,GHz with a fixed microwave power of $P_\text{rf}=0$\,dBm. The rectified microwave output $V_\text{D}$ provided by the diode is recorded with a voltmeter. The diode voltage $V_\text{D} \propto P_\text{rf}-P_\text{abs}$ is used as a measure for the field- and frequency-dependent microwave power absorption $P_\text{abs}$ in the device.

The BLS optical setup employs a continuous wave laser source with a wavelength of 532\,nm. The laser beam passes a polarizing beam splitter (PBS) and a $\lambda/2$-plate before it is focused onto the surface of the YIG film using a microscope objective lens with a focal length of 4\,mm (not shown in Fig.~\ref{setup}\,(a)). The focussed laser spot is positioned in the center of the SRR, where CST microwave studio\cite{CST} simulations of the unloaded SRR predict the most homogeneous microwave magnetic field. The incident laser photons are inelastically scattered by the magnonic excitations in the SRR/YIG film system. Hereby, the frequency of the inelastically scattered light is shifted by $\pm f$ in anti-Stokes (AS) and Stokes processes, respectively, where $f$ is the magnon frequency. The polarization of the inelastically scattered light is rotated by the scattering event by an angle $\beta$ with respect to the incident polarization direction.\cite{Cochran1988} In contrast, the elastically scattered light retains its incident energy and polarization.\cite{Demidov2004, Sebastian2015} All of the scattered and collected light passes again the $\lambda/2$-plate before it reaches the PBS. The PBS is then used to selectively direct the inelastically scattered photons (which underwent a polarization rotation) to a Tandem-Fabry-P\'{e}rot interferometer (TFP). 
The scattering cross-section of the magnons, and thus the polarization of the scattered light, is strongly dependent on the incident light polarization\cite{Cochran1988, LeGall1971}. The $\lambda/2$-plate allows to simultaneously rotate the polarization of the incident and backscattered light by changing the angle $\vartheta$ of its fast optical axis relative to the polarization axis of the incoming light. In combination with the PBS it is possible to analyze the polarization of the backscattered light with respect to the incoming light polarization.

Fig.~\ref{setup}\,(b) shows a typical BLS spectrum of the YIG film as function of the detuning $f$ from the laser line. At $f=0$\,GHz the photons from the elastic scattering process are observed while at $f=4.9$\,GHz photons inelastically scattered from the magnons by an anti-Stokes process are detected. The intensity $I$ of this peak is proportional to the number of magnons present in the system.\cite{Demokritov2001, Sandweg2011} To improve the signal-to-noise ratio, the integrated number of counts in the region of interest ($f\pm100$\,MHz) is used and denoted by $I_\text{AS}$.
\footnote{The frequency offset of the BLS spectra is corrected with respect to the microwave source.}

%\section{Experiment}
\begin{figure*}[t!]%
	\begin{center}
	\scalebox{1}{\includegraphics[width=0.9\linewidth,clip]{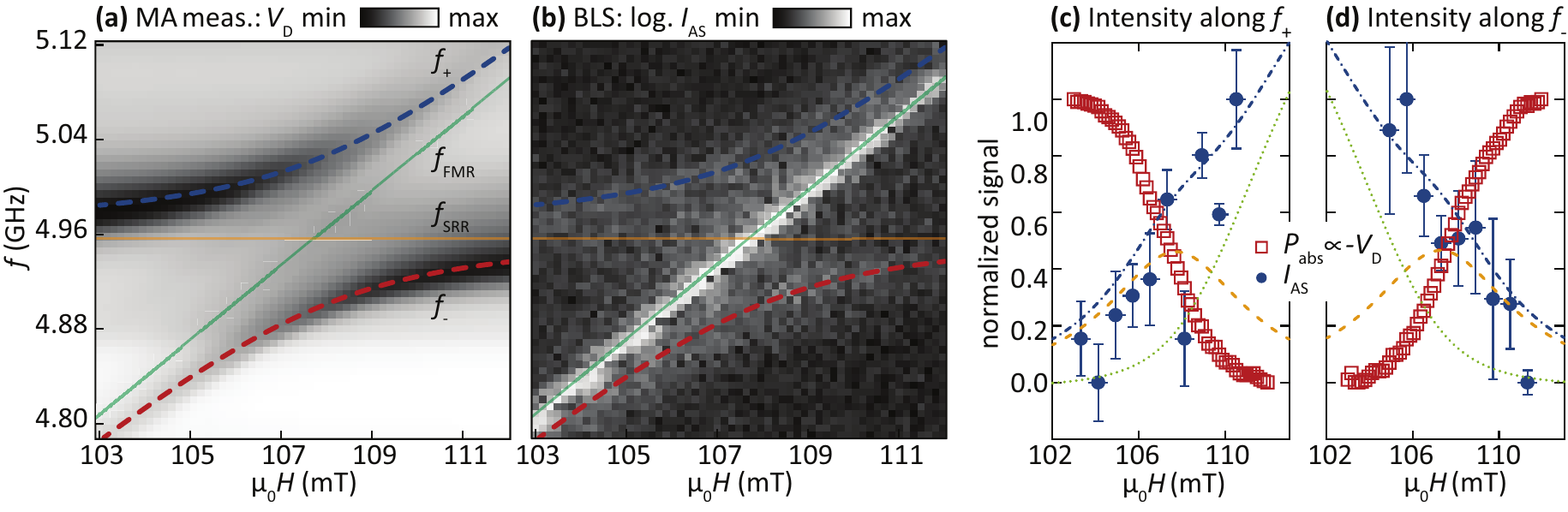}}
	\end{center}
	\caption{\label{colormaps} 
		(a) Microwave absorption versus applied magnetic field and microwave frequency. The anti-crossing phenomenon is clearly visible. (b) BLS anti-stokes intensity as a function of field and frequency. In addition to the anti-crossing, the uncoupled FMR mode of the YIG film is clearly visible. The dashed lines in (a) and (b) were obtained by simultaneously fitting the anti-crossing from (a) and the FMR mode from (b) to Eq.~\ref{coupling_equation}. The solid lines show the uncoupled modes.
		(c) and (d) Microwave absorption $P_{\rm abs}$ and anti-Stokes intensity $I_{\rm AS}$ from (a) and (b) along $f_+$ (c) and $f_-$ (d). For $f_+$ and $f_-$ close to $f_\text{FMR}$, strong AS intensity and weak MA is observed. Far away from $f_\text{FMR}$ the AS intensity vanishes and the MA becomes maximal. The lines indicate the contribution of the magnonic (orange dashed) and of the uncoupled FMR (green dotted) to the total BLS signal (blue dash-dotted).}%
\end{figure*}%

Firstly, the measurements without the $\lambda/2$-plate (corresponding to $\vartheta=0^\circ$) are discussed.
The YIG thin film and the SRR couple electromagnetically with an effective coupling strength $g_\text{eff}$. If $g_\text{eff}$ exceeds the intrinsic loss rates of the YIG and the SRR, an anti-crossing of the magnon and photon modes is expected.\cite{Wesenberg2009, Imamoglu2009,Huebl2013} To get information on both the photons in the SRR and the magnons in the YIG, the microwave absorption and the BLS signal are simultaneously recorded as function of the applied magnetic field and microwave frequency.
In Fig.~\ref{colormaps}\,(a) the microwave absorption is plotted versus the applied magnetic field and microwave frequency. At $\mu_0 H = 103$ and 112\,mT, a single strong resonance at $f_\text{SRR}$ occurs corresponding predominately to the pure microwave eigenmode of the loaded SRR.
For magnetic fields around about 108\,mT, where the detuning between the photon and magnon modes is zero, the mode coupling results in two hybridized modes with frequencies $f_+$ and $f_-$. Since the effective coupling strength $g_{\rm eff}$ is larger than the relevant loss rates, the mode coupling is observed as a pronounced anti-crossing. It is evident from Fig.~\ref{colormaps}\,(a) that the $f_+$ and $f_-$ modes approach the purely photon and magnon modes for large detuning. 

Fig.~\ref{colormaps}\,(b) shows the simultaneously recorded BLS signal. Here, three different modes are visible: (i) The most prominent mode appears at a frequency which depends about linearly on the applied magnetic field. This mode is attributed to the detection of the ferromagnetic resonance at frequency $f_{\rm FMR}$ excited directly by the feedline. \footnote{Depending on the position of the YIG film relative to the feedline the uncoupled mode can also be observed in the MW transmission as shown in Ref.~\onlinecite{Bhoi2014}}
(ii) The two faint modes at $f_+$ and~$f_-$ resemble the hybridized modes of the magnon-photon system. Their field and frequency dependence is identical to that of the hybridized modes detected in the microwave absorption experiments. The low intensity of these modes in the BLS measurement indicates a small BLS scattering cross-section. Note that the pure photon modes of the SRR are not detected by BLS. 

For a quantitative analysis of the hybridized mode frequencies $f_+$ and $f_-$ a model of two coupled harmonic oscillators is used:\cite{Bhoi2014, Huebl2013, Bai2015}
\begin{equation}
\label{coupling_equation}
f_\pm =  \frac{f_\text{SRR}+f_\text{FMR}}{2}\pm \sqrt{\left(\frac{f_\text{SRR}-f_\text{FMR}}{2}\right)^2 +\left(\frac{g_\text{eff}}{2 \pi}\right)^2}.
\end{equation}
Here, the SRR resonance frequency $f_\text{SRR}$ is assumed to be independent of the applied magnetic field. The ferromagnetic resonance frequency $f_\text{FMR}$ is modeled by the Kittel equation which describes the precession frequency of a macrospin in an in-plane magnetized ferromagnetic film:\cite{Kittel1948}
\begin{equation}
f_\text{FMR}=\frac{\gamma}{2 \pi} \mu_0 \sqrt{H\left(H+M_\text{eff}\right)}.
\end{equation}
Here, $\gamma=g\mu_\text{B}/\hbar$ is the gyromagnetic ratio, $g$ the Land\'{e} factor, $\mu_\text{B}$ the Bohr magneton, $\hbar$ the reduced Planck constant and $M_\text{eff}$ the effective magnetization of the YIG film. An excellent agreement of $f_\pm$ with the MA and BLS data is obtained, as can be seen by the dashed fit curves in Figs.~\ref{colormaps}\,(a),(b). From the fits $g_\text{eff}/ 2\pi=63\pm1$\,MHz, $\mu_0 M_\text{eff}=182\pm5$\,mT and $g=2.003\pm0.004$ are obtained. The extracted values of $\mu_0 M_\text{eff}$ and $g$ agree well to previously reported observations for similar YIG films.\cite{Klingler2014, Algra1982, Popov2012, Heinrich2011} The loss rate of the loaded resonator $\kappa/2\pi=25$\,MHz is determined from the half width at half maximum of the resonance at $\mu_0 H =40$\,mT, where the magnon and photon systems are well decoupled. The loss rate of the spin system $\eta$ is determined by moving the sample onto the feedline in order to record the FMR without the SRR. From this MA measurement $\eta/ 2\pi \leq18.5$\,MHz is found. Taken together, both $g_\text{eff}/\kappa > 1$ and $g_\text{eff}/\eta > 1$, and thus the system is well in the strong coupling regime.\cite{Zhang2014}

Fig.~\ref{colormaps}\,(c),(d) shows both $I_\text{AS}$ (full dots) and $ P_\text{abs}\propto-V_\text{D}$ (open squares) of the coupled system normalized to $[0,1]$ along $f_\pm$ in Figs.~\ref{colormaps}\,(a),(b). Note that $I_\text{AS}$ was averaged in a frequency and magnetic field range of $\pm3$\,MHz and $\pm0.25$\,mT, respectively,  to improve the signal-to-noise ratio of the BLS measurements. The standard deviation is shown by the error bars.
Fig.~\ref{colormaps}\,(c) shows a decrease of $P_\text{abs}$ along $f_+$ for increasing $H$, while $I_\text{AS}$ simultaneously increases. Along $f_-$, shown in Fig.~\ref{colormaps}\,(d), both $P_\text{abs}$ and $I_\text{AS}$ show the reversed trend. %In total a strong microwave absorption for a weak magnonic signal and vice versa can be found. 

In the following only the behavior along $f_+(H)$ is discussed for simplicity. The discussion of the $f_-$ mode is completely analogous. For small applied magnetic field mainly the photonic mode of the SRR is excited which leads to a high $P_\text{abs}$ and vanishing $I_\text{AS}$. For increasing magnetic field, $f_{\rm FMR}$ approaches $f_{\rm SRR}$  and the photon and magnon modes become increasingly hybridized. Since the photonic and magnonic character of the hybridized mode decreases and increases with increasing field, respectively, the same is expected for $P_{\rm abs}$ and $I_{\rm AS}$ probing the respective character of the hybridized mode.  At about 108\,mT ($f_{\rm FMR} =f_{\rm SRR}$) the photonic and magnonic character have equal weight and a drop of $P_{\rm abs}$ to 0.5 as well as an increase of $I_{\rm AS}$ to 0.5 is expected in good agreement with the experimental data. Increasing the applied field further reduces the photonic character of the $f_+$ mode and $P_{\rm abs}$ is expected to drop to zero for $\mu_0H\gg 108$ mT, where the $f_+$ mode is purely magnonic. Again, this is in good agreement with the experimental data of Fig.~\ref{colormaps}\,(c). Accordingly, $I_{\rm AS}$, which is probing the magnonic character of the $f_+$ mode, is expected to increase to unity. Although this also seems to be in good agreement with the data, the situation is more complicated here and has to be discussed in more detail. Due to the limited linewidth $\kappa$ of the SRR, the amplitude of the excited FMR mode is expected to rapidly decrease for $|f_{\rm FMR}-f| > \kappa/2 \pi$ and vanish for  $|f_{\rm FMR}-f| \gg \kappa / 2\pi$ as indicated by the dashed orange line in Fig.~\ref{colormaps}\,(c).\cite{Lotze2015} The resulting $I_{\rm AS}$ is expected to follow the orange line since it reflects the amplitude of the magnetic excitations.\cite{Buchmeier2007, Lotze2015}  The fact that this is not observed in the experiment is attributed to the excitation of the FMR mode by the microwave field from the feedline. The signal expected from this uncoulped mode is shown by the dotted green line in Fig.~\ref{colormaps}\,(c),(d). Evidently, the  dash-dotted blue line representing the sum of both contributions describes the measured BLS intensity reasonably well. 

\begin{figure}[t!]%
	\begin{center}%
		\scalebox{1}{\includegraphics[width=0.9\linewidth,clip]{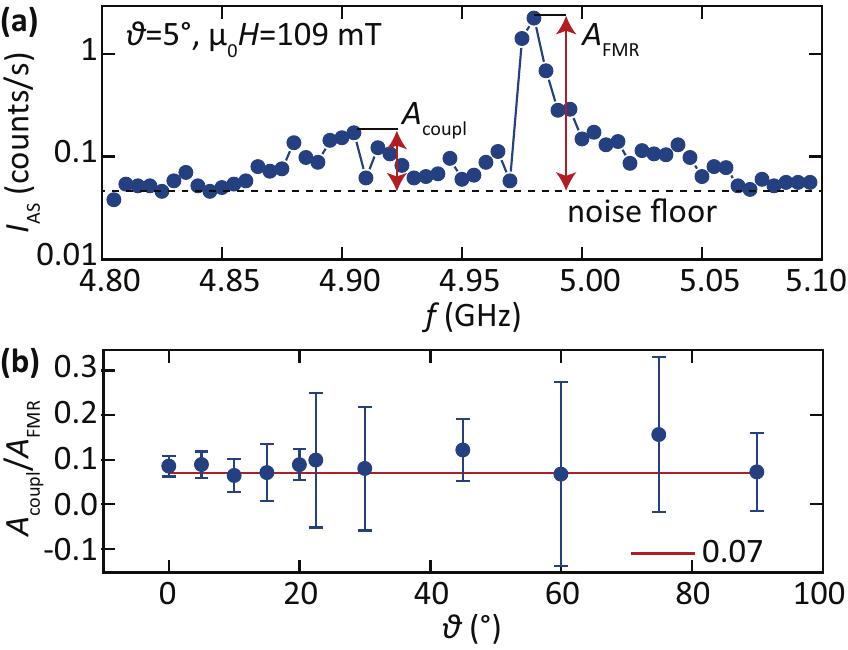}}%
	\end{center}%
	\caption{\label{polarization} 
		(a) Anti-stokes intensity plotted versus microwave frequency for $\theta=5^\circ$ and $\mu_0 H=109$\,mT. The noise floor is substracted to obtain $A_\text{coupl}$ and $A_\text{FMR}$. (b)~The ratio $A_\text{coupl}/A_\text{FMR}$ is plotted as a function of $\vartheta$. The red line marks the average ratio of 0.07. Within the error bars, $A_\text{coupl}/A_\text{FMR}$ is independent on $\vartheta$.
	}%
\end{figure}%

In a further set of experiments, the polarization of the inelastically scattered light is investigated by inserting the $\lambda/2$-plate into the optical path as shown in Fig.~\ref{setup}\,(a).
In Fig.~\ref{polarization}\,(a) an example $I_\text{AS}$ spectrum is shown for $\vartheta=5^\circ$ and $\mu_0 H=109$\,mT. Two peaks are visible in the spectrum: a weak one corresponding to the hybridized $f_-$-mode at $4.9$\,GHz and a strong peak stemming from the pure FMR mode at $4.98$\,GHz excited by the feedline. The amplitudes $A_\text{coupl}$ and $A_\text{FMR}$ of these two peaks are extracted relative to the noise floor as shown in Fig.~\ref{polarization}\,(a). Note that the difference $|f_--f_\text{FMR}|$ is sufficiently larger than the linewidth of both resonances, so that no influence of the pure FMR mode on the intensity of the hybridized $f_-$ mode is expected at this magnetic field value.

Fig.~\ref{polarization}\,(b) shows the ratio $A_\text{coupl}/A_\text{FMR}$ as a function of $\vartheta$ for constant $\mu_0 H=109$\,mT. The error bars depict the uncertainty in the determination of the noise floor. The average value of $A_\text{coupl}/A_\text{FMR}\approx0.07$ is shown by the red line.
Within error bars, $A_\text{coupl}/A_\text{FMR}$ is independent of $\vartheta$. This is consistent with the notion that all photons inelastically scattered off the hybridized modes and the pure FMR mode undergo the same polarization rotation. The findings indicate that only the magnonic part of the hybridized mode is accessible in the BLS measurement. The photonic part does not change the BLS process, suggesting a vanishing photon-photon scattering probability.

In conclusion, the presented microwave absorption and BLS measurements reveal strong magnon-photon coupling and its up-conversion to optical frequencies in a system consisting of a micropatterned split-ring resonator and a YIG film. A coupling constant of $g_\text{eff}/ 2\pi=63$\,MHz is found, which exceeds the loss rates of both the pure spin system and the split-ring resonator. 
The combined analysis of the microwave absorption and BLS intensities strongly indicate a continuous transition from a photonic to a magnonic mode with varying applied microwave frequency and magnetic field. The measurements show that the BLS and MA techniques are complimentary by sensing the magnonic and photonic character of the hybridized excitations, respectively.

The experiments presented here provide a powerful platform for the study of time-dependent oscillations of the coupled system in between the purely magnonic and photonic states during coherent magnon-photon exchange, as well as limiting decoherence processes. Furthermore, the BLS technique opens the path to study magnon Bose-Einstein condensates\cite{Demokritov2006} coupled to a photonic resonator and the coherent exchange between magnonic supercurrents\cite{Clausen2015} and photonic resonators.

During the preparation of the manuscript we became aware of similar experiments by Nakamura and coworkers.\cite{Hisatomi2016, Osada2016}

Financial support from the DFG via SPP 1538 ''Spin Caloric Transport``, Project GO 944/4 is gratefully acknowledged.

\end{document}